\RequirePackage{fix-cm}
\documentclass[smallextended]{svjour3}       
\smartqed  
\usepackage{csquotes}
\usepackage{xcolor}
\usepackage{graphicx}
\usepackage[numbers]{natbib}
\usepackage[hyphens]{url}

\begin{document}

\title{A Pragmatic Approach to Regulating\\Artificial Intelligence:}
\subtitle{A Technology Regulator's Perspective}


\titlerunning{A Pragmatic Approach to Regulating Artificial Intelligence}        

\newcommand{\todo}[1]{{\color{red}#1}}

\author{Joshua Ellul\and
        Stephen McCarthy\and
        Trevor Sammut\and
        Juanita Brockdorff\and
        Matthew Scerri\and
        Gordon J. Pace
}


\institute{Joshua Ellul (corresponding author) is Chairperson of the Malta Digital Innovation Authority and Director of the Centre for Distributed Ledger Technologies at the University of Malta.
\at \email{joshua.ellul@um.edu.mt}
\and
Stephen McCarthy is the inaugural CEO of the Malta Digital Innovation Authority.
\at \email{stephen.mccarthy@mdia.gov.mt}
\and
Trevor Sammut is the Chief Regulatory Officer of the Malta Digital Innovation Authority.
\at \email{trevor.sammut@mdia.gov.mt}
\and
Juanita Brockdorff is a Partner at KPMG in Malta.
\at \email{JuanitaBrockdorff@kpmg.com.mt}
\and
Matthew Scerri is Associate Director of Digital Solutions at KPMG in Malta.
\at \email{MatthewScerri@kpmg.com.mt}
\and
Gordon J. Pace is a Professor of Computer Science with the University of Malta.
\at \email{gordon.pace@um.edu.mt}
}

\date{Uploaded: 15th April 2021}

\maketitle

\begin{abstract}
Artificial Intelligence (AI) and the regulation thereof is a topic that is increasingly being discussed within various fora. Various proposals have been made in literature for defining regulatory bodies and/or related regulation. In this paper, we present a pragmatic approach for providing a technology assurance regulatory framework. To the best knowledge of the authors this work presents the first national AI technology assurance legal and regulatory framework that has been implemented by a national authority empowered through law to do so. In aim of both providing assurances where required and not stifling innovation yet supporting it, herein it is proposed that such regulation should not be mandated for all AI-based systems and that rather it should primarily provide a voluntary framework and only be mandated in sectors and activities where required and as deemed necessary by other authorities for regulated and critical areas.

\end{abstract}

\section{Introduction}
\label{sec:intro}
Issues concerning the design of legal and regulatory frameworks for Artificial Intelligence (AI) have been a topic of discussion and debate for the past few decades. Much of the debate inherits from discussions on regarding how to regulate technology and the regulation of computer systems, but reaches further due to the very nature of the potential for AI. In fact, one can argue that a substantial portion of the debate is due to this very potential, which brings together ethical issues, rights, perils and other aspects. Regardless of which school of thought one subscribes to, in a spectrum ranging from the requirement of generic principles \cite{asimov2004robot}, to specific laws\footnote{\url{https://www.europarl.europa.eu/doceo/document/E-9-2019-002411_EN.html}}, to advocating that regulation of such technology should be avoided, and focus should be on safety mechanisms~\cite{yampolskiy2013artificial}, when the technology is used for applications that can directly or indirectly impact society then sufficient regulation (whether through law or otherwise) should be investigated (whether applied directly or indirectly to the technology). At the same time whilst some argue for mandatory regulation, many warn that regulation could stifle innovation \cite{gurkaynak2016stifling}.


In this paper, we do not purport to present a contribution to this philosophical debate, but rather our aim is a more pragmatic one --- that of outlining and explaining the rationale behind a legal and regulatory framework addressing AI systems adopted by Malta. Whilst other regulatory frameworks and bodies have been proposed in literature (discussed in Section~\ref{sec:rel}), it is to the best knowledge of the authors that the framework being presented herein is the first AI technology assurance legal and regulatory framework that has been implemented by a national authority (the Malta Digital Innovation Authority\footnote{https://mdia.gov.mt}) empowered through law \cite{malta2018mdia} to do so. 

Towards the end of the 2010's Malta built a framework for addressing the regulation of Innovative Technology Arrangements \cite{malta2018itas}, in order to ensure better end user protection through the adoption of appropriate due diligence on the underlying technologies. Initially focusing on Blockchain, Smart Contracts and other Distributed Ledger Technologies (DLTs)~\cite{ellul2020regulating}, the legislation has since been extended to cover critical systems (through a legal notice\footnote{\url{https://legislation.mt/eli/ln/2020/389/eng/pdf}}) and regulatory guidelines have been issued by the Malta Digital Innovation Authority (MDIA) for the regulation of arrangements which use an element of AI. 

The aim of the paper is to provide a review of the regulatory framework proposed and to put it in the context of the ongoing AI regulation debate. One of the primary observations is that the need for precise definitions and objective measures in the legal framework, meant that the Maltese regulatory approach is founded on practical and auditable aspects, and is intended to address concerns with existing technology (as opposed to attempting to address possible issues arising from future development of AI technology, for example Artificial General Intelligence). To implement an AI regulatory framework intended for modern day technology and also in aim of not stifling innovation, the framework is primarily voluntary however may be mandated based upon the sector and/or risk associated with the activity within which the AI system is used or as deemed necessary by another lead authority or governing legislation. This sets the tone of much of the paper, but it is naturally endemic to any discussion of practical implementations of the regulation of technologies. The fast evolving nature of technology requires law-makers to address existing technology in a sound manner, but also in a way that is expected to be future-proof.

The rest of the paper is structured as follows. Next, in Section~\ref{sec:caseforass}, we build a case as to why technology assurances and associated frameworks are necessary, and then in Section~\ref{sec:voluntary} further make a case as to why such frameworks should primarily be voluntary for AI-based systems yet should be mandatory for AI-systems used within regulated and critical activities. In Section~\ref{sec:aiita} we introduce an AI innovative technology arrangement regulatory framework. We put our work in context of related work in Section~\ref{sec:rel}. Then in Section~\ref{sec:conc} conclude with closing thoughts and future directions.

\section{The Case for AI Assurances}
\label{sec:caseforass}
In this section, we highlight issues relating to AI-based systems which could result in systems operating incorrectly in relation to its intended functionality and thereafter build the case for instilling assurances. We concentrate on Artificial Narrow Intelligence (ANI) given that the state-of-the-art has not yet reached levels of Artificial General Intelligence (AGI) \cite{bostrom2014ethics}. We will use the term AI throughout the rest of the paper to refer to AI that exists today --- ANI. 

Since the inception of software development, the fact that such systems occasionally fail has been accepted to be the norm. Although much work has gone into developing techniques to reduce the frequency and severity of such occurrences, we continue to experience software malfunction on a daily basis. The impact of such failure is contained as long as the software functions in a closed system i.e. it has no direct impact on the real world\footnote{One can argue that no software system is closed. For instance, a bug in a game marketed as pure entertainment will still impact the value the user obtains from the game with respect to the price payed, and a bug crashing an image editing program may still lead to loss of productive hours put in by an employee.}, but frequently software affects the real world in a direct or indirect manner\footnote{An early example of indirect impact can be found in Babbage's Analytical Engine, an endeavour which was largely driven by his interest in computing nautical tables efficiently. Direct impact can vary in criticality --- from the software controlling a light system, to plane auto-pilot software.}. 

One finds reports of many catastrophic failures in literature and newspaper reports. Some of which led to huge financial losses, with one of the highest-profile cases being that of the financial services company Knight Capital Group which, in 2012, endured a loss of 460 million US dollars in less than an hour due to a bug in a trading algorithm~\cite{knight}. Others have led to operational failure, such as a bug in electricity generation and distribution monitoring software which led to the electricity load across a number of US and Canadian states not being redistributed resulting in an hours-long blackout in 2003~\cite{electricity}. Finally, some failures have led to actual or potential loss of human life, such as a case in 2014, where Carefusion had to recall over 50,000 infusion pumps due to a bug which could have led to drugs being infused earlier or later than intended with possibly lethal results~\cite{carefusion}.

Various reasons are often cited why traditional engineering practice has failed to eradicate bugs in software systems, with the most common being that the complexity of software systems far exceeds that of most traditionally engineered systems. However, software engineering has led to the development of various techniques to enable the identification and hence removal of critical bugs --- from testing techniques which explore a system's state space through a number of potential environmental conditions to static verification techniques which attempt to explore the state space more thoroughly (though at an additional verification cost in terms of time and human expertise required), to runtime verification and fault tolerance techniques in which additional code is used to capture the symptoms of bugs while a system is running and attempt to act in order to avoid their effects. 

AI systems are no exception when it comes to incorrect behaviour, even when the algorithms themselves are correctly implemented incorrect behaviour could emerge as will now be discussed. For a particular learning-based AI task, consider that the implementation of the AI software does not have any bugs. This means that the functionality of the AI algorithm will operate correctly --- it will learn how to make decisions based upon the training information fed into it and experience it undergoes and will act accordingly. This does not necessarily mean that the system will function correctly in relation to the expected behaviour though. Such incorrect behaviour can be attributed to insufficient training, biases and unbalances that may exist within datasets, and due to limited AI models which may require more levels of complexity e.g. consider an AI vision algorithm intended to detect different types of species of dogs, which does not first check whether the image is that of a dog --- the algorithm may incorrectly classify a cat as a particular dog species. 

Undoubtedly AI systems should undergo standard quality assurances processes. The AI algorithm should be tested not only for functional correctness of the algorithm itself but also with respect to the behaviour emergent following training. Expected system behaviour should be checked and validated for a sufficient amount of data and use cases. However, like traditional software testing, testing of AI systems is only as good as the coverage of training data, iterations and permutations and use cases which are undertaken. Once an AI system is deployed and it encounters an event that it was not trained to handle it may well end up handling it incorrectly. More so, if it is continuously learning in a live environment it may be exposed to certain situations which could affect its behaviour negatively.

Verification methods have been investigated for certain types of knowledge-based systems \cite{tsai1999verification}. However, when it comes to using static verification for learning-based systems, they may provide insight in regards to aspects of correctness of the implementation, but not of the expected runtime behaviour, which is likely to be intractable given the state explosion of both static analysis and learning algorithms. 

Runtime verification techniques for ensuring that certain behaviour is abided by (and other unwanted behaviour is not) has been used in the context of autonomous systems \cite{goldberg2005runtime} and can provide a level of assurance in regards to behaviour execution bounds. This provides assurances that irrespective of what a system has learnt, it will continue to operate within expected behaviour bounds, and if it does not remedial action can be taken. We later discuss a similar generic concept of a harness in Section~\ref{sec:aiita}.

Many AI-based techniques are like black-boxes. For such techniques, we can understand the code and the algorithm's mechanics, however to a large extent (for many types of algorithms) can not know what they have learnt. Therefore, we may not be able to predict the outcome of decisions they may take and/or we may not be able to provide justification in regards to why it had taken certain decisions. It is for this reason that extensive research towards explainable AI is being investigated. In which some advocate that we should only use AI techniques for which its decisions can be explained, whilst others are working to provide methods to explain decisions made by AI algorithms which are currently deemed to not be explainable.

The past decade has seen various infamous cases where unexpected behaviour emerged from AI systems, sometimes of a controversial or even safety-critical nature. Take for example Google's controversial classification algorithm that classified some people as gorillas \cite{miller2018od}; or IBM's cancellation of the `Watson for Oncology' project, with some allegations that the tool provided useless, occasionally dangerous recommendations \cite{strickland2019ibm}.

The increasing concern is not only to do with cases that have emerged but also based on the reality that more and more systems are becoming computerised and automated. One often-referenced cliché is that of automated scoring systems (for example insurance issuance that may introduce elements of discrimination based on demographics) \cite{kirkpatrick2016battling}. Indeed, such discrimination is not ideal, and therefore efforts should be undertaken to ensure any bias in datasets is removed and attempts made to remove discriminatory features during training. That being said the algorithm will always be fit to the model and data which it has been exposed to, and will be biased to that data. The challenge is to ensure that the bias towards the particular training data is not representative of any biases that would be undesirable.

The concerns highlighted above demonstrate the need to ensure that sufficient assurances are put in place to ensure that AI algorithms are implemented correctly, and that their behaviour is as expected and does not introduce any unwanted biases. Indeed, many are advocating for such regulatory frameworks to be developed and applied to AI systems. However, need such frameworks be mandatory for all AI-based systems? We now follow with a case for why such frameworks should not always be mandated, and that they should not be focused on the technology but should be focused on the sector or activity that the AI is being used within/for.

\section{The Case for Voluntary Assurances of AI and Mandatory Assurances of Regulated and Critical Activities}
\label{sec:voluntary}
Above we introduced why various effort should be applied to ensure that AI systems function correctly both in terms of algorithmic correctness, as well as in terms of system behaviour, and also to ensure that systems do not introduce aspects of bias and any other unwanted behaviour. 

Many have proposed mandatory regulation of AI. Again, ignoring AGI, when it comes to ANI should such frameworks always be mandated? Consider the case of an AI movie recommendation system. If the system were to be used for personal use only, trained using the individual's preferences and model trained using only features pertaining to movies (e.g. genre, actors, producers, narratives, sequence of events and their timings, etc). Should such a system really require to be regulated? Now consider, that the system is extended to include other consumers' preferences. Besides aspects of data privacy and protection (which should be factored in), should the AI algorithms be regulated? Should the AI's recommendations be seen any different to a critic's recommendations? Depending upon cultures one may state that movie themes could be an issue of controversy. Ignoring such issues, or perhaps instead considering a system for an activity where there are clearly no cultural or controversial issues such as a system that learns your sleeping patterns in order to save energy around your house, is there really a need to mandate regulation to ensure that no issues of bias are introduced? Actually, bias in this particular case would likely be welcomed.

Now consider that the underlying software framework used for different AI systems is the same, and the difference between implemented systems is how consumers decide to wire them up to real-world applications. In this case, the underlying infrastructure is agnostic to the application above. Should such underlying infrastructure be required to be regulated? Also, if regulation is mandated for AI systems, where does one draw the line with respect to what is AI and what is not AI? Often technologists can make clear personal distinctions, however, it becomes increasingly hard to define this in an objective manner. More so, what difference does it make if an algorithm is AI-based or not and yet can be used for the same activity? Then, should we be talking about AI regulation at all? Or should we be focusing on software? Or rather the activity it is used for irrespective of how it is implemented (even through human operators).

If all AI would need to be regulated and/or follow guidelines then this may shackle and stifle innovation \cite{gurkaynak2016stifling}. More so, just the definition of AI alone is controversial, and even if a definition is chosen, is it going to be clear what software is AI and what software is not? There are some algorithms which we can ascertain are universally accepted as AI, and some systems which are universally considered to not have aspects of AI within them, however what should be done about the rest? Could this approach not only stifle AI-innovation, but also other software based innovation?

Looking back at the principles of regulation though, we need to ask ourselves why is regulation of AI being proposed? Is it only because of end-of-the-world scenarios being painted which require AGI, which the state-of-the-art is currently not capable of? If so, then perhaps we should differentiate between any regulatory requirements for AGI and ANI. We propose that this should be done, at least in the interim until AGI is deemed to be upon us. We leave considerations for AGI as future work, and here will continue discussing aspects pertaining to ANI.

If regulation is required for AI for non-regulated and non-critical activities, given the lines between AI, software in general and automation are blurred, and given that non-AI based techniques and even manual human processes could end up taking the same `incorrect' decisions that such AI would take, then why should we not regulate all software and all manual human processes? Well, because software, automation, AI and manual processes are used to conduct all ranges of activities that do not require any form of regulation governing them. Therefore, herein we propose that stakeholders should focus on mandatory regulation of sectors/activities and not of technology, to exemplify consider the following:

\begin{itemize}
  \item Take an insurance scoring example. Whether AI, a non-AI system or a human makes a decision to issue or not an agreement or decide on risk levels, the process should be required to ensure that there is no bias. Then mandatory insurance regulation can dictate such requirements and it is up to the operator to ensure that such processes do not have bias.
  \item In the case of surveillance equipment, using the same argument above, we should not be implementing mandatory AI/software regulation, but laws should be enacted with respect to how one can/cannot undertake surveillance.
  \item Finally, consider a compliance engine built into a payment service, which decides which transactions should be allowed to go through. Whether or not built using AI, the crucial element is that transaction filtering should abide by the laws regulating financial services.
\end{itemize}

To amplify this argument consider credit scoring. Discrimination has been a problem since the early days of (non-software based) manual credit scoring systems \cite{capon1982credit}. In 1974, US Congress passed an Equal Credit Opportunity Act to ensure individuals were not discriminated against on the basis of gender or marital status. Then further, in 1976 race, religion, age and other characteristics were also included within the act. This act is focused on the activity, i.e. credit scoring, and not the technology used to help with credit scoring. For such anti-discriminatory laws still in place today, the laws must be abided by irrespective of the technology used.  Different activities may or may not impose similar or different anti-discriminatory laws or none at all. It does not make sense to duplicate specific activity focused laws within the context of technology, and therefore in this particular case we should steer clear of mandating AI laws in this regard. The specific activity's laws (say for example insurance laws) should specify that such discrimination is not introduced in the activity, and an associated regulator may monitor and enforce the law as deemed necessary. The respective activity's laws and regulator could then decide upon the ideal way forward with respect to automated processes (that may or may not use AI). For example, they could mandate that operators need to ensure that no such discrimination takes place and (perhaps) that decisions made must be explainable and demonstrate that no such discrimination formed part of each particular decision irrespective of whether AI, software, some other technology, or manual processes were used.

The question of what constitutes high-risk or defines whether a sector or activity should mandate this framework arises. This is left up to other lead authorities and laws of the land to decide. For financial affairs, a financial services authority (a separate body) may impose when a sector or activity should be mandated to undertake a technology audit (as proposed herein), or even if any levels of enhanced due diligence is required. In fact, the Malta Financial Services Authority\footnote{http://mfsa.mt} mandates MDIA audits for operations within the cryptocurrency sector. Similarly, a health authority may require that applications deemed to be of a critical nature do the same (e.g. the Malta national health authority mandated an MDIA system audit for the national COVID-19 tracing app implemented). However, one might ask how to decide on whether to mandate the same for applications that may at surface level be benign yet may in fact have strong implications and tangible effects on society --- for example large scale social media platforms and their associated recommendation systems \cite{schmitt2018counter}. The question at hand is not really about the technology (AI), but is about the activity (social media). Whether laws could be created to cater for this situation, or a body empowered to oversee such a sector (which could very well be entrusted to a technology regulator), or otherwise, the work herein does not provide answers in regards to how to go about this question. Yet it provides a mechanism in regards to how one aspect (technology assurances) of regulating such a sector can be achieved if it is deemed necessary.

Therefore, based on the above we make the argument that mandatory regulatory frameworks should not be technology-specific (or AI-specific), yet should be activity or sector-specific as defined and required per activity/sector (i.e. insurance, banking, housing, welfare, etc) and/or other lead authorities or laws. However, such non-technology based regulators may find challenges to ensure that systems abide by their required regulation and also that the systems are implemented correctly (both functionally and behaviourally). Therefore, we opine that such lead authorities should mandate where required that technology assurance regulatory frameworks are abided by to ensure that such AI algorithms are implemented correctly, that runtime behaviour is as expected (as well as testing for any discriminatory features that should not exist).

AI technology-based assurances may not only be required for regulated activities, however various AI-based products and services may see the benefit to provide assurances to various stakeholders that their systems have been verified to be of a sound nature with respect to functional and behavioural correctness. Therefore, the regulatory approach enables for technology-based assurances to also be offered on a voluntary basis (besides being mandated from lead authorities of respective sectors/activities).

Now, we present the AI technology assurance framework implemented by the Malta Digital Innovation Authority\footnote{\url{https://mdia.gov.mt}} which offers certification of AI systems on a voluntary basis where sought, or on a mandatory basis where other lead authorities or laws require it. We believe that such a balance of mandatory regulation where required and voluntary where sought is a regulatory balance which both ensures that technology used is properly kept in check, and yet at the same time does not stifle innovation by requiring mandatory regulation where it is not required.

\section{An AI Technology Assurance Regulatory Framework}
\label{sec:aiita}

We now present the AI Innovative Technology Arrangement (AI-ITA) regulatory technology assurance framework. Approaches for providing software assurances will invariably have a degree of commonality irrespective of the technology domain and also application domain within which the solution is categorised under. As such, this framework builds on the Innovative Technology Arrangement (ITA) \cite{ellul2020regulating} regulatory assurance framework overseen by the Malta Digital Innovation Authority (MDIA). Rather than mandating compliance and certification of all AI based systems, the regulatory framework is a voluntary one --- unless a lead authority deems that such technology assurances are required. For example, the financial authority may mandate that automated AI trading platforms would require certification, and similarly the transport authority may require such mandatory compliance for driverless cars. It is in this manner we believe innovation can still flourish, by only requiring mandatory oversight of sectors and activities that should require such oversight.

\subsection{AI Innovative Technology Arrangement}
The challenge with Artificial Intelligence ITAs (AI-ITAs), primarily revolves around the wider categorisation of what constitutes AI. This necessitates a deeper level of understanding of what a particular technology arrangement is and does, with a wider consideration to the different risks and mitigation mechanisms required. The subjectivity in what is considered AI also presents a unique challenge, that of being able to define what is an AI-ITA, and thus eligible for certification.

Rather than define what is an AI-ITA as a hard and fast rule, the guidelines take the approach of defining qualities and criteria that qualify software as an AI-ITA. Other than the criteria that the logic of the software must be based on underlying datasets, one or more of the following would be required:

\begin{enumerate}
	\item The ability to use knowledge acquired in a flexible manner in order to perform specific tasks and/or reach specific goals;
	\item Evolution, adaptation and/or production of results based on interpreting and processing data;
	\item A systems logic based on the process of knowledge acquisition, learning, reasoning, problem solving, and/or planning; and
	\item Prediction forecast and/or approximation of results for inputs that were not previously encountered.
\end{enumerate}

The above ensures that techniques and algorithms commonly associated with the wider AI field are captured and include anything from Deep Learning to Natural Language Processing and Optimisation Algorithms. The MDIA will also continue to monitor developments and update guidelines as required to include (and potentially exclude) defining features of what is/not classifed as an AI-ITA.

\subsection{System Audits, Auditors and Subject Matter Experts}
The framework provides a structure for the Authority and applicant to work with independent (and approved) system auditors to be able to scrutinise to a fairly high level of detail the software itself as well as the manner with which it is being operated under the ISAE 3000 \cite{iaasb2013isae} standard for assurance. The audit of the software system itself is primarily conducted via a code review, whose aim is to ensure that the manner with which the AI-ITA is implemented accurately reflects what the organisation behind the AI-ITA are claiming in their technology blueprints. The rationale behind this is to ensure that any claims being done are truly reflected in the code, which enables the general public, who may not know what AI really is to gain trust in the system given that it stood up to scrutiny prior to the certificate being issued. Beyond the software, the certification mandates depending on the type of audit being undertaken and associated controls, to also give the general public assurances that the AI-ITA creator and operator are running the organisation in a manner that meets the standards set out by the MDIA. The certification therefore enables the general public to trust the creator, in the manner they build, maintain and run the AI system. Two main types of audits are required throughout an AI-ITA's lifetime: (i) first a `Type 1 Systems Audit' is required which focuses on providing assurances with respect to functional correctness typically undertaken as an AI-ITA's first audit; and (ii) a `Type 2 Systems Audit' which focuses on renewing previous assurances provided through a previous audit which factors in live data and operations associated with the system to assure the system assurances are still in place within the period under audit.

The audit process begins with the applicant submitting a request (in the form of an application) to the Authority, upon which the Authority will assess the applicant by reviewing the provided documentation around the AI-ITA and conduct its due diligence. Following this, the MDIA issues a Letter of Intent upon which the applicant will be able to appoint an MDIA approved Systems Auditor, and notify the MDIA of the appointment, for the MDIA to verify that the Systems Auditor has the required competencies (which the Authority has tested the system auditor for). The Systems Auditor will then conduct the audit as per the Authority's guidelines\footnote{\url{https://mdia.gov.mt/wp-content/uploads/2019/10/AI-ITA-Guidelines-03OCT19.pdf}} and compile a report with their findings, which is issued to the MDIA for a review and a subsequent decision on whether the certificate is to be issued. Once issued, a further follow up audit must be conducted every time there is a material change in the AI-ITA (and on renewal after every two years).

Systems Audits are an integral part of the certification process as they provide the MDIA with an independent report on the particularities of the AI-ITA, specifically the code and whether it accurately reflects what is being disclosed in the blueprint, and the ongoing operations of the AI-ITA. Systems Audits are conducted by Systems Auditors, who must be independent from the AI-ITA and its operator, that are subject to approval by the Authority, and who need to meet a set of requirements (defined in the Systems Auditor guidelines) through their combined complement of Subject Matter Experts (SMEs) in the fields of IT audit, cybersecurity and technology with specialisation, in this case, in AI. The SMEs will be the primary individuals responsible for conducting systems audits, and must adhere to a set of requirements, such as ensuring that they meet a level of continuous professional education in the AI field\footnote{\url{https://mdia.gov.mt/sa-guidelines/}}.

This section describes the requirements that an AI-ITA must meet in order to qualify for certification. The end-goal of these requirements are for the MDIA to be able to certify the system, the manner with which it is developed and how it is operated. Following we will introduce salient features of the regulatory framework.

\subsection{ITA Blueprint} 
The Blueprint document is an essential document in the certification process as it is meant to provide a detailed description to the Authority on what the system does, how it's designed, and operated. Other than allowing the MDIA to evaluate whether AI-ITA certification is applicable, and to assess the initial levels of quality of a particular AI-ITA, it is further intended to be used by the Systems Auditors as the document against which aspects such as the code is reviewed against. The blueprint also defines a minimum set of disclosures that must be disclosed to direct users (in English) in a non-technical manner, to be able to communicate the features and functionalities of the system and how it respects the ethical AI framework\footnote{\url{https://malta.ai/wp-content/uploads/2019/08/Malta_Towards_Ethical_and_Trustworthy_AI.pdf}}, limitations to prevention of bias, and the expected accuracy of the AI-ITA.

In a general (AI agnostic) sense, the detailed description must cover the reasons for which the system was created, the functional capabilities of the AI-ITA, how the system is to be verified and tested to ensure the results meet expectations and what the operational limitations of the systems are. More specifically, for an AI solution the blueprint must include a disclosure of the AI techniques used and to justify why certification is being sought, and how (depending on the technique that applies) specific risks are being managed and mitigated --- which could include, for example, what is being done to ensure that the underlying dataset does not have discriminatory biases within it. In a broader sense, the Blueprint must highlight the safety mechanisms in place and alignment with Malta's Ethical AI Framework. The Blueprint guidelines are accompanied by a Technology Stack Nomenclature document\footnote{\url{https://mdia.gov.mt/wp-content/uploads/2019/10/AI-ITA-Nomenclature-03OCT19.pdf}}, which delves into further technical detail and explains how the information must be disclosed. Crucially it defines a process map showing how the information may be disclosed in order to ensure that the manner with which the system works is adequately disclosed.

\subsection{ITA Harness} 
A crucial element that the AI-ITA framework proposes, and which needs to be highlighted clearly in the Blueprint is the ITA Harness. The ITA Harness provides a safety net for the process by monitoring activity inputs and outputs to ensure that the boundaries (which must also be disclosed in the Blueprint) are respected. Furthermore, the ITA harness must also be able to handle any anomalies it detects (such as outputs outside expected boundaries) in a manner which is also disclosed. The AI-ITA harness must also communicate with the Forensic Node (discussed next) to ensure that any anomalies are appropriately logged and can be investigated and rectified. While the harness may not apply to all AI-ITAs, the Authority requires that when it does not apply it must be justified adequately in the blueprint and accompanied with alternative plans of how the behaviour of the AI-ITA will be monitored and contained\footnote{\url{https://mdia.gov.mt/wp-content/uploads/2019/10/AI-ITA-Blueprint-Guidelines-03OCT19.pdf}}.

\subsection{Forensic Node}
The Forensic Node is another requirement mandated by the MDIA, and whose implementation and operation is also subject to the audit. The purpose of the Forensic Node is to ``store all relevant information on the runtime behaviour of the AI-ITA in real-time such as recording of inputs and outputs, and supporting data related to potential explainability of how an output was derived from a given input wherever applicable''. This means that any inputs, outputs as well as data that supports how the system achieved the results it did must be stored in a secure data store in real-time. This highlights that the Forensic Node is not only used to support the assessment of (some of) the operating effectiveness of the controls during an audit, but may also be used to support legal compliance by the MDIA (or other authorities) and also enables a further layer of monitoring to be done (manually or automated) by a Technical Administrator (discussed next). It is important to note that the Forensic Node must be separate from the ITA Harness, in that the Forensic Node is more concerned with Data Logging, as opposed to the monitoring in relation to boundaries. Ultimately the data in the Forensic Node may also assist the AI-ITA operator in conducting analysis of the AI-ITAs historic behaviour, particularly when deploying AI systems that change and adapt over time based on the information provided to them (such as evolutionary algorithms or reinforced learning techniques).

\subsection{Technical Administrator}\label{sec:ta}
A Technical Administrator, a form of a service provider appointed by the AI-ITA to act as the final safeguard for the system, must be appointed and in place at all times. The Technical Administrator must be able to intervene, if required to do so by the MDIA, another authority or legally (such as in the event of a breach of law by the AI-ITA), to limit further impact to the users and where necessary limit or reverse losses. For example, consider an AI system that utilises reinforced learning and which, after a period of time, starts to exhibit discriminatory bias that goes against the principles laid down in the ethical AI framework and/or against the requirements of any laws or rules it must abide by. In this case the Technical Administrator must be able to halt the operation of the system to prevent further damage and revert to an older model (as may be mandated by a legal judgement). As such, this also imposes an indirect requirement for the AI-ITA to provide mechanisms to enable the Technical Administrator to conduct their actions as may be necessary (e.g. by ensuring regular snapshots of the machine learning models are kept to revert back to).

\subsection{English Description and Consumer Protection}
The system being certified is checked by the systems auditors who, amongst other things, ensure that its functionality matches that described in the blueprint in human-readable form (in English). If, post-deployment, the system exhibits behaviour contrary to this description against which it was certified, the Innovative Technology Arrangements and Services Act specifies that the English version prevails legally.

\subsection{Auditing of Design and Development Processes}
Systems Audits include oversight of the design and development process of the system-under-audit. Not only does such oversight cover traditional software engineering principles, but for systems including an element of AI also includes assurances that certain foundational principles have been taken into consideration in the process.

\noindent\textit{Build on a human-centric approach.} The systems auditors ensure that the AI system was designed in a manner to support and assist humans without overriding the user into taking any unwanted decisions and the manner with which it operates musts be equitable and inclusive across different segments of society.

\noindent\textit{Adherence to applicable laws and regulations.} It is crucial that behaviour induced by the system, including parts driven by AI, will be designed in a manner that adheres to the law.   

\noindent\textit{Maximise benefits of AI systems while preventing and minimising their risks.} It is crucial that any risks induced through the use of AI are identified and mitigated accordingly, including the setting up of controls to ensure fairness, transparency and resiliency to new AI-specific attack vectors. 

\noindent\textit{Aligned with emerging international standards and norms around AI ethics.} As the world is increasingly becoming globalised through technology, and which may be further amplified through the proliferation of AI systems, this objective was laid down to ensure that Malta’s ethical framework is aligned with similar ethical guidelines by the EU commission\footnote{\url{https://ec.europa.eu/digital-single-market/en/news/ethics-guidelines-trustworthy-ai}} and OECD\footnote{\url{https://www.oecd.org/going-digital/ai/principles/}}.

The framework further builds on these principles by delineating a number of principles (such as Human Autonomy, Fairness, Prevention of Harm and Explicability) and proposes 63 controls of how these can be tested. While not all of these controls apply to all AI-ITAs, the AI-ITA must show that it has taken them into consideration and justify in the Blueprint (and ultimately top the users) those controls which do not apply.




\section{Related Work}
\label{sec:rel}
The work presented herein is related, complimentary and orthogonal to a number of different areas which we will now provide an overview of including proposals for setting up AI focused regulatory bodies, AI and software audit standards, continuous monitoring and algorithmic compliance and explainability, other non-technology assurance related areas of AI law and regulation, and ethical frameworks. 

\subsection{Towards Regulatory Bodies}
The European Parliament had proposed for the setting up of an ``European Agency for robotics and artificial intelligence in order to provide the technical, ethical and regulatory expertise needed to support the relevant public actors'' \cite{delvaux2016motion}. Setting up an EU body specifically for AI may cause overlap with other existing bodies (including ENISA and DG Connect) \cite{fitsilis2019development}. 

The sentiment to create such a regulator was further proposed emanating from the need for transparency of automated systems handling personal data and the often impossibility of doing so due to trade secret protection \cite{wachter2017right}. A solution proposed was to allow for a trusted third party to undertake an audit of the system in question \cite{wachter2017right}. It was further proposed to develop an European regulator to undertake such audits \cite{wachter2017right}. On similar lines, to avoid differing domestic approaches (which could result in regulatory competition \cite{smuha2019race}), the need for an International Artificial Intelligence Organisation was highlighted \cite{erdelyi2018regulating}. Indeed, this would be a step in the right direction, however it is the opinion of the authors that the need for providing regulation should not wait for such an organisation to emerge, yet national authorities (such as the MDIA) could work together towards harmonisation and adapt to eventual international standards and guidance as it emerges.

To minimise overlap amongst regulatory bodies and make best use of limited available expertise, Malta opted to set up a regulatory body (the MDIA being discussed herein) responsible for innovative technologies that go beyond AI unlike that proposed by \cite{delvaux2016motion}, whilst also opting for a local national body rather than wait for a supra-national body to be formed (as proposed by Delvaux \cite{delvaux2016motion} and Wachter \cite{wachter2017right}). Fitsilis proposed a `hierarchy of regulation'  \cite{fitsilis2019development} for the `chain of multi-level governance' including global, supra-national and national regulatory bodies. Such global and supra-national regulatory bodies did not (and do not) exist and therefore national regulation was developed independent of (non-existent) supra-national and global requirements. That said, as standards and regulation may be developed higher up in the chain the local national regulatory framework could make changes in aim of global harmonisation. 

The Maltese approach adopted in 2019 has been noted by the EU \cite{european2020white} which ends with a proposal where the Commission recommends that Europe equips itself to undertake testing and certification of high risk AI systems. The Commission differentiates between `high-risk' and `non-high-risk' AI applications, taking the position that only the former should be in the scope of a future EU regulatory framework. In line with the Maltese approach, at a harmonised EU level `low-risk' AI applications would be governed by a voluntary labelling scheme. As regards compliance and enforcement, the EU Commission considers prior conformity assessments, which could include `procedures for testing, inspection or certification' and/or `checks of the algorithms and of the data sets used in the development phase' – in the same vein as the MDIA's certification procedure. 

Benefits and drawbacks can emanate from various forms of regulation proposed from government controlled regulation and enforcement to self-regulation (and also no regulation) \cite{clarke2019regulatory}. To support a robust yet flexible regulatory framework, differing levels of regulation had been applied and developed in a co-regulatory manner as similarly suggested by Clarke \cite{clarke2019regulatory} in aim of ensuring the regulatory framework is proportionate and conducive to the various stakeholders.

A proposal for an Artificial Intelligence Development Act (AIDA) \cite{scherer2016regulating} aimed to start a conversation in regards to how to best proceed to regulate AI in such a manner that innovation is not stifled yet at the same time associated risks are lowered. This work proposes that a specific authority is set up to oversee and regulate the space. This work was academic in nature, whilst the Malta Digital Innovation Authority (MDIA) and Act are first real-world instances of such. Contrary to the proposed AI-specific authority though, the MDIA's remit encapsulates innovative technologies and software beyond just AI. Such a decision is crucial as an AI-specific authority and/or legislation would either leave loopholes allowing for similar types of software (falling outside of a definition of AI) to escape regulatory requirements, or would capture all software related activities that would ideally be best not included within such a regime (and would also create further potential overlap with other potential technology regulators). Thereafter, it is the authors' opinion that AI-specific guidelines and guidance could be provided.

\subsection{International Standards}
Whilst global software regulatory bodies do not (yet) exist, global standards do. The International Organisation for Standardization (ISO) has developed a number of different standards for use within the software domain\footnote{\url{https://www.iso.org/ics/35.080/x/}}. Whilst, such software focused standards can be useful for global recognition within the framework described herein local national standards were required to be developed for the following reasons: (i) standards available to date do not provide guidelines or comprehensive control objectives specific for the artificial intelligence domain; (ii) mechanisms and roles for ensuring continuous monitoring and intervention are not defined \cite{coallier1994iso, boiral2011managing} (within which the MDIA has incorporated requirements for a forensic node and technical administrator); and (iii) the authority is ultimately responsible and empowered through the MDIA Act to ensure audit integrity and quality whilst at the same time able to propose changes to legislation and guidelines in light of the sector's nascency. Once international standards adequate to adopt are developed national guidelines may be updated to make use of them (if deemed to meet the national requirements). That said, the authority has adopted and requires that audits are undertaken following the ISAE 3000 \cite{iaasb2013isae} standard which specifies generic (i.e. not software nor AI related) principles for quality management, ethical behaviour and performance for use in non-financial areas.

\subsection{Algorithmic, Computational and Continuous Monitoring and Explainability}
Audits in general are known to increase cost for regulators and to be burdensome for businesses and operators \cite{hulstijn2011continuous}. To minimise costs and burden for the various involved stakeholders algorithmic and computational auditing and compliance have been proposed \cite{raji2020closing, wang2018regulatory}. Computational and algorithmic audits can be seen as, rather than an alternative to human-controlled audits, two sides of the same coin that compliment each other to attain higher levels of assurances with respect to a technology arrangement's correctness and adequacy. Such work compliments the proposed framework in that system auditors may utilise and/or deem computational and algorithmic auditing processes in place as adequate measures to fulfil requirements of various control objectives. The authority empowers system auditors to use software and their judgement in regards to whether control objectives are met. Such a design decision was important to ensure that the framework was as flexible as possible in regards to the types of software arrangements it can support. That said, in future, computational auditing for specific use-cases or sectors could be supported by the authority by building infrastructure required which could be utilised by operators.

Continuous monitoring can, not only further help decrease the costs of audits, however can detect incidents or deviations ``at or near real time, before they escalate and turn into a real problem'' \cite{ezzamouri2018continuous}. Again, it is not a case of choosing between a traditional audit or auditing with continuous monitoring software or processes, since they are complimentary. The framework proposed herein requires for `Type 2 System Audits', i.e. system audits the are focused on the correct continual operation of a technology arrangement, to be undertaken. Indeed, such audits are performed at an instance in time rather than involve continual monitoring, however a system auditor may make use of continuous monitoring software or processes to deem that at the time of a Type 2 System Audit, various control objectives were met based upon information retrieved via continuous monitoring tools. Furthermore, the framework requires that a forensic node is set up which will be collecting any information of relevance to the correct or incorrect operation of the technology arrangement. As described in Section \ref{sec:ta} a technology administrator should both ensure the correct operation of the forensic node, and intervene if any alerts are raised. Therefore, though the definition of a forensic node is very open, to ensure that the regulatory framework is flexible, it can be seen that the forensic node in many instances will in fact be implementing a form of continuous monitoring.

Based on the above (and other proposals), mechanisms for automated compliance can be put in place. This would further ease the burden and costs (beyond initial costs) on operators and regulators. In aim of putting into place a regulatory framework that could aide in raising the levels of technological assurances sooner rather than later, such a mechanism was not included within the first instance of this regulatory framework. To enable for such a future a number of challenges would need to be overcome including: (i) rather than rely on system auditors to interpret control objectives and ensure that they have been met, concrete objective computational rules must be encoded in such a way that they it would be possible to check for their compliance at run-time; and (ii) current control objectives define what behaviours must be checked for, but not how such behaviours are neither implemented nor checked \cite{vazquez2008human}. Furthermore, provision of such guidelines and/or systems enabling automated compliance could further support legal compliance through design enabling a ``more appropriate, ethical and responsible response to complex compliance requirements'' \cite{hashmi2018legal}.

On a similar note extensive work is being undertaken on the explainability of AI systems (both from technology and law perspectives \cite{atkinson2020explanation}). Whilst, ultimately, it is the role of system auditors to ascertain that functionality, behaviours and qualities of a system are verified, aspects pertaining to trustworthiness of a system could be built through explainability \cite{GrzegorzExplain} --- interestingly the relationship between explainable systems and system auditors draws parallels to that of AI and Law, which at the heart ``points to common subject matter that underlies the two fields: the coordination of human behaviour'' \cite{bench2012history}. As the field of explainability matures, future work on providing guidance with respect to its use within the auditing domain should be investigated.

\subsection{Other Non-technology Assurance Related Aspects}
Extensive research has been undertaken on other aspects pertaining to technology regulation and legislation (which are beyond the focus of this paper). A number of challenges that exist within this space (which should be covered in future work) include how to handle issues of liability \cite{balkin2015path, barfield2018liability, rachum2019whose, scherer2016regulating}, intellectual property and copyright \cite{gurkaynak2016stifling, gamito2021algorithmic}, sector specific regulation (e.g. regulation of AI-based health systems \cite{o2019legal}, autonomous vehicles \cite{gurkaynak2016stifling}, finance \cite{wischmeyer2020regulating}, social media \cite{wischmeyer2020regulating} and its use within law \cite{rissland2003ai, bench2020need}), privacy and data protection \cite{gamito2021algorithmic, ebers2019regulating, wischmeyer2020regulating, gurkaynak2016stifling}, fundamental rights, profiling and anti-discrimination issues \cite{gamito2021algorithmic, goodman2017european, reed2018should, buiten2019towards}, competition law \cite{wischmeyer2020regulating}, and legal personality \cite{wischmeyer2020regulating}.

Quite a number of ethical frameworks have been proposed \cite{floridi2018ai4people, ebers2019regulating, wischmeyer2020regulating} and ``at least 63 public-private initiatives have produced statements describing high-level principles, values and other tenets to guide the ethical development, deployment and governance of AI'' \cite{raji2020closing, mittelstadt2019ai}. Within the framework proposed herein an ethical framework is referenced to, however the scope of such a discussion would warrant a paper of its own.

\section{Conclusions}
\label{sec:conc}
In this paper we have highlighted the need for an AI technology assurance regulatory framework which is implemented in a manner that both promotes technology and does not stifle innovation, yet at the same time enforces assurances where required. This was proposed through a voluntary regulatory technology assurance regime, which can be mandated where required for specific sectors or activities (e.g. for health, finance, safety-critical infrastructure, etc) by other lead authorities or laws. 

In this paper we have presented a world first implementation of an AI national regulatory framework that is overseen by a national technology regulator (the MDIA). Based on discussions above we now present a number of recommendations for various jurisdictions and AI stakeholders:
\begin{itemize}
    \item Mandatory regulation of AI should be avoided, due to its potential ability to stifle innovation. In fact, the issues at hand are not necessarily about AI, but extend to software and automation in general. More so, the issues being discussed (e.g. discrimination) are not centered on technology, but are concerned with the processes put in place for particular activities --- irrespective of whether the process is undertaken by a human or software (or hardware). Therefore, sector/activity focused regulation should impose where required relevant restrictions and requirements. For example, rather than impose that all AI should ensure that it does not discriminate, specific laws on activities should be mandated --- e.g. insurance providers may not discriminate against applicants and must be able to provide clear explanations in regards to application decision processes (again irrespective of whether the decision was made using or by AI, other software, a human or a hybrid thereof).
    \item AI is and will continue to be increasingly used in all aspects of life and sectors. Expertise in AI (and other innovative technologies) are scarce. More so, technology is cross-cutting and more often than not technology related assurances and oversight are not sector specific. Therefore, we recommend that other jurisdictions investigate into setting up a national technology regulator to put in to place such a voluntary regulatory framework which other national regulators (e.g. finance, health, communications, etc) can then work together with to require mandatory technology assurance regulation where required. This will ensure both best use of national resources as well as consistency across sectors in regards to technology assurances and processes.
\end{itemize}

Furthermore, some of the issues described above are also relevant to other innovative technologies such as:
\begin{itemize}
    \item Blockchain, Distributed Ledger Technologies (DLT) and Smart Contracts in which assuring correct functionality of their implementation may be crucial due to their immutable nature and potential inability to stop them. Then, consider the implications of an unstoppable AI agent (based on DLT). The issues discussed above are largely amplified.
    \item Quantum computing which will enable for exponentially more powerful computation, which could potentially threaten security of systems and the internet at large --- until infrastructure is upgraded to be able to stop potential attacks.
    \item The vision of a world of Internet of Things (IoT) where our everyday lives will be supported through the use of pervasive and ubiquitous computing devices embedded into various environments and aspects of our lives, where incorrect functionality or operation of such devices could cause critical or event catastrophic events (especially when used in critical infrastructure).
\end{itemize}

Moving towards a more efficient digital world has its benefits, however it brings various risks that need to be mitigated through adequate levels of technology assurances, and therefore such regulatory oversight should be put in place where required, and limit regulation (or stipulate no regulation) when such technology is used for non-critical applications and sectors and/or where no legal or ethical issues arise.

\bibliographystyle{spbasic}      
\bibliography{refs}   

\end{document}